\documentclass[useAMS,usenatbib, usegraphicx]{mn2e}

\usepackage{array}

\title[first experimental demonstration of temporal hypertelescopes]{First experimental demonstration of temporal hypertelescope operation with a laboratory prototype}
\author[L. Bouyeron, S. Olivier and Other]{L. Bouyeron$^{1}$, S. Olivier$^{1}$, L. Delage$^{1}$\
, F. Reynaud$^{1}$\thanks{E-mail:
francois.reynaud@xlim.fr}\newauthor P.Armand$^{2}$, E. Bousquet$^{2}$, J. Benoist$^{2}$.\\
$^{1}$ XLIM, dept. Photonique/IRO,\\
$^{2}$ XLIM, dept. DMI/IRO, 123 rue A Thomas,87060 Limoges CEDEX, France}
\begin{document}

\date{17 09 09 ...}

\pagerange{\pageref{firstpage}--\pageref{lastpage}} \pubyear{2009}

\maketitle

\label{firstpage}

\begin{abstract}
In this paper, we report the first experimental demonstration of a Temporal HyperTelescope (THT).  Our breadboard including 8 telescopes is firstly tested in a manual cophasing configuration on a 1D object. The Point Spread Function (PSF) is measured and exhibits a dynamics in the range of 300. A quantitative analysis of the potential biases demonstrates that this limitation is related to the residual phase fluctuation on each interferometric arm. Secondly, an unbalanced binary star is imaged demonstrating the imaging capability of THT. In addition, 2D PSF is recorded even if the telescope array is not optimized for this purpose.
\end{abstract}


\section{Introduction and context}
\subsection{Hypertelescope concept}
High resolution Optical imaging instruments based on aperture synthesis have been developed over the last decades with the aim of reaching angular resolution in the nano radian range. These different instruments \citep{Law} use the property of the Zernike van Cittert theorem to recover the intensity distribution of the object by means of spatial coherence analysis. With this method, the instrument can never select the light coming only from one of the pixels composing the full object because the measuremnts are being carried out on the Fourier spectral domain. For high-dynamics objects, such as  a star + exoplanet system, this technique is limited since the information on the faint object is always mixed with the light emitted by the main source. Consequently, direct imaging is to be preferred and the analysis of the object is made easier in the image domain than in the Fourier spectrum one. Since the beginning of high resolution imaging, measurements have never been achieved both with a very high resolution in the range of nanoradian and a very high dynamics in the range of $10^6$. In order to meet this challenge A. Labeyrie has proposed a solution which is known as the hypertelescope \citep{L}. This new type of instrument solves the problem of the highly structured Point Spread Function (PSF) of a diluted array thanks to a pupil densification process. The PSF of a hypertelescope being sharp and smooth, it is possible to use the instrument for direct imaging. The image $I$ which equals the convolution of the object $O$ by the PSF, looks like the object but with a limited resolution. Different versions of hypertelescopes have been proposed using field combination in the pupil plane \citep{V} or pupil densification thanks to the use of monomode optical fibres \citep{P}.
\subsection{Principle of classical Hypertelescope}
\begin{figure}
   \centering
   \centering
   \includegraphics[width=8cm]{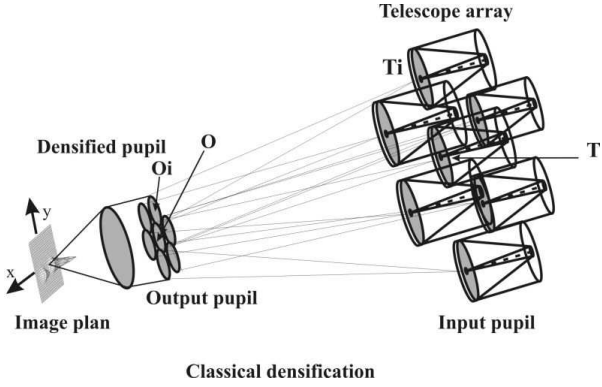}
      \caption{Spatial hypertelescope : the input pupil is reconfigured
      with a G contraction ratio and a densification of the output
      pupil $(TT_i =G\cdot OO_i)$. The image is displayed as a function of spatial
      coordinates $(x;y)$ in the image plane.
              }
         \label{Fig:SpatHyp}
   \end{figure}

Parallel to the hypertelescope studies promoted by A. Labeyrie, we have proposed a temporal alternative to the initial design that used spatial classical optics (cf Fig.\ref{Fig:SpatHyp}). The main purpose of this new concept is to answer some technical difficulties met with classical hypertelecopes and to propose new functionalities. \\
In \citet{RD} we theoretically demonstrated the possibility to design a hypertelescope by using temporal optical path modulation. In the next paragraphs,we briefly recall the principle of a classical and a temporal hypertelescope.

%

Figure \ref{Fig:SpatHyp} recalls the structure of a hypertelescope as proposed by A. Labeyrie. This simplified drawing does not
detail the reconfiguration and densification process. This technique makes it necessary to remap the input
pupil taking care to apply an homothetic contraction of beams center  distribution in the output pupil according to the "golden rule of imaging interferometry". Denoting $1/G$ the pupil reconfiguration ratio, the distance between telescope $T_{i}$ and the input pupil center $T$ is related to the position $OO_{i}$ of the sub pupil $i$ in the output
pupil by:

 \begin{equation}
     OO_{i}= \frac{TT_{i}} {G} 
   \end{equation}
where O is the output pupil origin.
After passing through the focusing lens, the beam $i$ in the image plane generates a limited plane wave:
	\begin{equation}
      E_{i} = a_{i} \cdot \exp{(j\psi_{i})} \cdot d(X) \cdot \exp{(jK_{i} \cdot X)}
      \label{field}
   \end{equation}

   with

 \begin{equation}
      K_{i} = \frac{2\pi}{\lambda}\frac{OO_{i}}{f} =
      \frac{2\pi}{\lambda}\frac{TT_{i}}{Gf}
   \end{equation}

where $f$ is the focal length of the focusing lens $L$, $a_{i}$  and  $\Psi_{i}$ are respectively the modulus and input phase of
beam $i$. $X$ is the position vector in the image plane with two coordinates $x$ and $y$ ($X = (x;y)$ ).
 In the Labeyrie configuration $d(X)$ denotes the field envelop resulting from the diffraction of each sub pupil supposed to be identical for all beams.
 The last term $\exp{(jK_{i} \cdot X)}$ describes the linear variation of the phase in the image plane with $ \alpha_i$ and $ \beta_i$ slope
coefficients along $x$ and $y$ axes.

 \begin{equation}
      K_{i} = 2\pi (\alpha_i , \beta_i)
   \end{equation}
   with
 \begin{equation}
      \alpha_i = \frac{(OO_i)_x}{\lambda f} \\
      and \\
      \beta_i = \frac{(OO_i)_y}{\lambda f}
   \end{equation}
where $(OO_i)_x$ and $(OO_i)_y$ represents the projection of  vector along $x$ and $y$ axis.

This way, each position $TT_i$ of telescope drives the slope $(\alpha_i,\beta_i)$ of the corresponding wave front and reaches the image plane. The phase evolution $\varphi_i(x, y) = K_iX$  can be analysed along different axes parallel to the $x$ axis. The phase variation  $\varphi_i(x)$ is linear with an offset depending on the $y$ coordinate :

 \begin{equation}
      \varphi_i(x) = 2\pi \cdot( \alpha_i \cdot x + \beta_i \cdot
      y) = 2\pi \cdot \alpha_i \cdot x + \varphi_{i0}(y)
   \end{equation}
with
 \begin{equation}
\varphi_{i0}(y) = \varphi_{i0}(0; y) = 2\pi \cdot \beta_i \cdot y
\label{phase_equa_temporal}
   \end{equation}

As demonstrated by A. Labeyrie, the possibility to retrieve the image results from the coherent superposition of these different
limited plane waves and the convolution of the resulting Point Spread Function (PSF) with the object intensity distribution. 

The PSF is given by the intensity distribution corresponding to the coherent superposition of field $E_i$ in the image plane when the
input pupil is illuminated by a plane wave with a wave vector perpendicular to the input pupil plane. Under such conditions, the
$a_i$ coefficients are identical (set to one for example) and the $\Psi_i$ phase terms in the equation \ref{field} are equal to 0.

 \begin{equation}
PSF(X) = \Bigg| \sum_{i}{E_i}\Bigg|^{2} = 
\Bigg|d(X)\Bigg|^{2}\cdot \Bigg| \sum_{i} \exp{(jK_i\cdot
X)}\Bigg|^{2}
   \end{equation}

If in the Labeyrie configuration,  $d(X)$ corresponds to the sub pupil far field, in the IRAN (Interferometric Remapped Array Nulling) proposal this term is replaced by the
sub-pupil function itself. In both cases this envelope term fixes the phase modulation span in the recombining
plane. This phase range is directly related to the optical pieces dimension and will not be easily adjusted in a real experiment.

For a tilted point like source illuminating the telescope array with a $\theta = (\theta_x, \theta_y)$ obliquity, the amplitudes
$a_i$ remain constant but the phase $\psi_{i}$ (as defined in eq (\ref{field})) becomes:

 \begin{equation}
\psi_{i} = \frac{2\pi}{\lambda}TT_i \cdot \theta
   \end{equation}

The total phase of the $E_i$ field can be written in the image plane as :

 \begin{equation}
\psi_{i} + \varphi_i = \frac{2\pi}{\lambda}TT_i \cdot \theta +
\frac{2\pi}{\lambda} \frac{TT_i}{Gf}\cdot X
   \end{equation}

This additional $\psi_{i}$ term induces a spatial shift of the corresponding intensity distribution :

 \begin{equation}
I_\theta(X) = PSF(X+Gf\theta)
   \end{equation}

For an extended object, the incoherent superposition of the different contribution of the object $O(\theta)$ leads to the fallowing image intensity distribution :

 \begin{equation}
I(X) \propto \int O(\theta) \cdot PSF(X+Gf\theta)d\theta
   \end{equation}

Of course, any process able to achieve such phase modulations will provide an image with the same basic properties. The next paragraph
deals with the possibility to obtain such a result in the time domain.

\subsection{From spatial to temporal domain}

As above mentioned, in a spatial hypertelescopes, the linear phase modulations $\varphi_i(x) $ are related to the $\alpha_i$ and $\beta_i$ tilts of the optical fields $E_i$ reaching the observation plane.

\begin{figure}
   \centering
   \includegraphics[width=6cm]{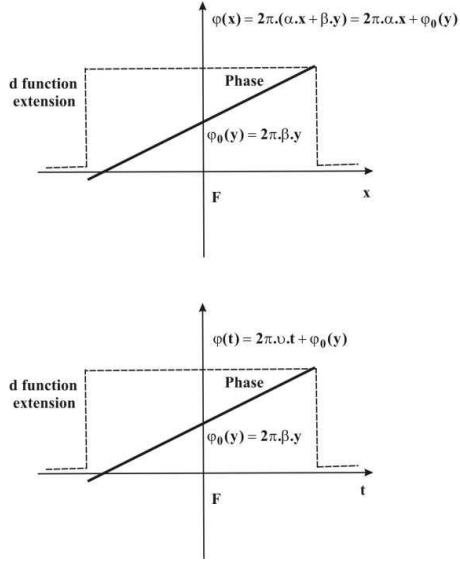}
      \caption{Phase variation in the spatial and temporal cases : In the spatial case (top), the phase variation is induced by the obliquity of plane wave related to each aperture. In this case, the phase span is limited by the diffraction spot size.
      In the temporal case (bottom), a phase modulation is applied by the operator to temporally reproduced the slope observed in spatial case. The phase span is now driven by the optical path stroke and can be adjusted without any limitation.}
      \label{Fig:Phvar}
   \end{figure}

In the temporal case, the phase modulation $\varphi_i(t)$ is generated thanks to an optical path variation induced by optical
path modulators linearly actuated as a function of time as shown on figure \ref{Fig:Phvar}.

 \begin{equation}
\varphi_i (t) = 2\pi \cdot \nu_i \cdot t + \varphi_{i0}(y)
\label {phase_equa}
   \end{equation}

   Using a temporal frequency $\nu_i$ scaled on the $\alpha_i$ slope allows to propose a temporal configuration equivalent to the
spatial concept. For this purpose, the frequency $\nu_i$ has to be  proportional to $\alpha_i$ with an arbitrary $A = \frac{\nu_i}{\alpha_i}$ coefficient.

 \begin{equation}
\nu_i = A \frac{(OO_i)_x}{\lambda \cdot f} =  A
\frac{(TT_i)_x}{\lambda \cdot f \cdot G}
   \end{equation}

According to this spatial to temporal transposition, the optical field:

 \begin{equation}
E_{i} = a_{i} \cdot \exp{(j\psi_{i})} \cdot d(X) \cdot
\exp{\{j[2\pi \cdot \alpha_i \cdot x + \varphi_{i0}(y)]\}}
   \end{equation}

is replaced by :
 \begin{equation}
E_{i} = a_{i} \cdot \exp{(j\psi_{i})} \cdot d(t) \cdot
\exp{\{j[2\pi \cdot \nu_i  \cdot t + \varphi_{i0}(y)]\}}
   \end{equation}
$d(t)$ function expresses the span of phase variation. In the classical concept, the image is analyzed along the
$x$ axis. The $A$ factor represents the relationship between the spatial parameter $x$ and time $t$.
 \begin{equation}
x = A \cdot t
   \end{equation}

%
%
%

In the time domain, the image is obtained by using the temporal PSF
given by:

 \begin{equation}
PSF(t; y) = \Bigg| \sum_{i}
d(t)\cdot \exp{\{j[2\pi \cdot \nu_i \cdot t + \varphi_{i0}(y)
]\}}\Bigg|^{2}
   \end{equation}

As seen in the previous case, for a tilted point like source with a $\theta = (\theta_x
;\theta_y)$ obliquity, the total phase of the $E_i$ field becomes:

 \begin{equation}
\psi_{i} + \varphi_i = \frac{2\pi}{\lambda}TT_i \cdot \theta +
\frac{2\pi}{\lambda} \frac{TT_i}{Gf}\cdot (At; y)
   \end{equation}
The corresponding intensity distribution is a shifted temporal PSF:
 \begin{equation}
I_\theta(At; y) = PSF[(At; y)+Gf\theta]
   \end{equation}
 and the incoherent superposition of the different contribution of the object $O$ leads to the temporal image:
 \begin{equation}
I(At; y) \propto \int O(\theta) \cdot PSF[(At; y)+Gf\theta]d\theta
   \end{equation}

\begin{figure}
   \centering
   \includegraphics[width=6cm]{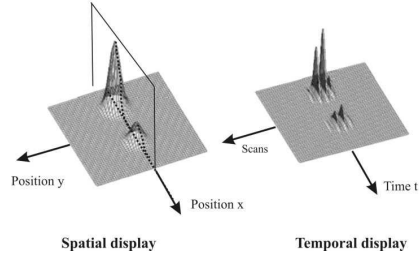}
      \caption{Temporal versus spatial display. The two-dimensional
      spatial display can be replaced by a temporal one. As long as
      time is a one dimensional parameter a set of scan is necessary
      to retrieve a time multiplexed image like in a video monitor.
              }
         \label{Fig:Tempdisp}
   \end{figure}

This result demonstrates the full equivalence between spatial and
temporal display as illustrated in figure \ref{Fig:Tempdisp}.

%
%
%

In order to analyze a two dimensional image, the $\varphi_{i0}(y)$ phase has to be modified step by step in order to display successively the
different s of the image using a set of parallel scans. The information provided along the $x$ axis at a given position $y$ is
fully equivalent to the one observed as a function of time $t$ for the scan related to the same $\varphi_{i0}(y)$ value.
 The image is temporally multiplexed as for a video raster scan. It can be noticed that in the spatial domain, the $x$ span is
determined by the diffraction field distribution related to the sub pupil geometry. In the temporal domain, the image field is determined by the extension of phase modulation directly driven by the span of the optical path modulator. It can be adjusted very easily and reduced down to $0$ contrary to the
spatial configuration in which the limitation results from the beam dimensions. The different advantages of temporal hypertelescopes are discussed in \cite{RD}.

\section{Implementing a temporal Hypertelescope}

\subsection{General layout}
\begin{figure}
   \centering
   \includegraphics[width=8cm]{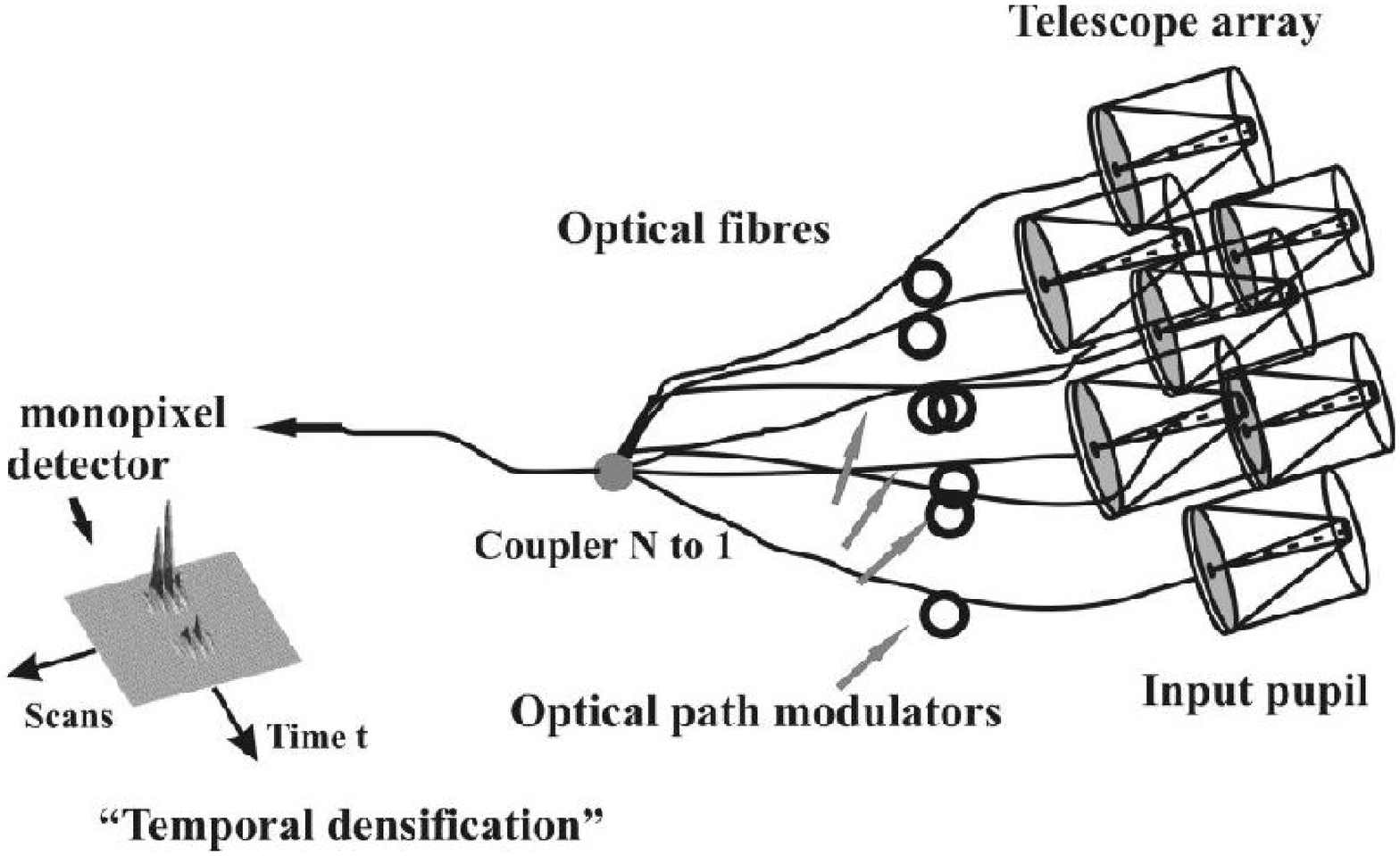}
      \caption{Proposal for a temporal hypertelescope using
      optical fibres and a guided coupler acting as a combiner.
      The optical path modulators induce the phase variations
      necessary to form an image. The image is temporally multiplexed.
              }
         \label{Fig:Temppropal}
   \end{figure}

Figure \ref{Fig:Temppropal} shows a generic representation of a temporal hypertelescope. This sketch uses optical fibres and an optical coupler but classical mirror trains, air delay lines and classical beam splitters could be used to design such an instrument.\\ 
Let's consider a telescope array pointing at a scientific target. The light is picked up at the N telescope foci and passes through optical path modulators before being recombined
using an N to one beam combining coupler in a coaxial configuration. The required optical path modulations of the N interferometric arms are generated using
optical fibre stretchers. They induce an optical path variation $\delta_i$ in order to generate the convenient phase modulations as previously
mentioned. For this purpose, the optical path modulation can be expressed as :

\begin{equation}
\delta_i  = \frac{A(OO_i)_x}{f} \cdot t +
\frac{(OO_i)_y \cdot y}{f}
   \end{equation}

These optical path modulations can be servo controlled using opto-electronic systems  previously developed for such kinds of applications
(\cite{D}; \cite{O2}). It allows to monitor, with a nanometric accuracy, the linearity of the optical path variation
as a function of time. For each scan, the offset $\frac{(OO_i)_y \cdot y}{f}$ is set in order to display the signal that would
be observed at $y$ position for a classical spatial hypertelescope.

%
%
%

To properly operate a temporal hypertelescope, the optical path modulators are driven by an 8 channel function generator and related high voltage electronics (not drawn in the picture). The output voltage drives the optical path modulators with a full span in the range of tens $\mu m$ and a typical nanometric  sensitivity. The electronic gain and the voltage generator slopes allow to set the $\nu _i$ frequencies at the proper values. 
In such a configuration, we can theoretically get the same imaging properties as for the first classical design using spatial pupil densification. 
The breadboard described in this paper and the related experimental results reported in the next paragraphs aim to demonstrate the validity of this new concept.

\begin{figure*}
	\begin{center}
		\includegraphics{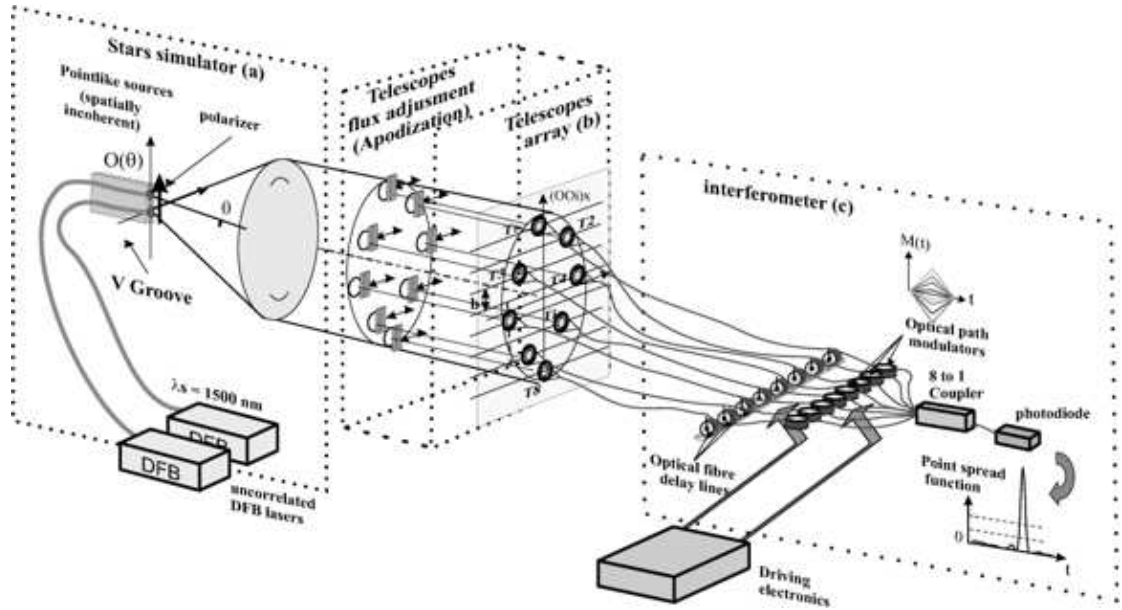}
		\caption{General scheme of our Temporal hypertelescope (THT) test bench. The main sub systems are described in detail in the text: The star simulator, the telescope array and the combining interferometer.}
		\label{schema_complet}
	\end{center}
\end{figure*}

\subsection{Technological options}
Our experimental set-up  (see figure \ref{schema_complet}) has been designed and implemented thanks to the different skills developed in our team for two decades (\citealt{All}; \citealt{S}; \citealt{H}; \citealt{Perrin}; \citealt{O}). Consequently, our THT experimental test bench uses optical fibres and couplers for the different  optical functions to be implemented.  However, we would like to stress that the use of guided optics components is not mandatory for the implementation of a THT. A classical design with classical components could be chosen if preferred. This point will remain a minor one as long as we will focus more on the demonstration of THT principle than on the technological aspects. 
The following items give the general framework of our experimental study.
\begin{itemize}
\item{The operating wavelength, all over this instrument, is $\lambda=1,55\mu m$ to take advantage of the mature and available technologies of Telecom components. }
\item{Light propagation between the entrance pupil and the detector is achieved through monomode polarization maintaining fibres (panda fibres).}
\item{All connections between the different components use FC/APC connectors to avoid parasitic back reflections.}
\item{The THT configuration has been optimized for the imaging of an unbalanced binary star with a high dynamics such as an exoplanet-star couple. For this purpose we will use the theoretical results reported in \citet{A} which implies a redundant spacing of the array and an optimized power distribution over the different telescopes to apodize the point spread function.}
\item{Our telescope array includes 8 apertures equivalent to a linear redundant configuration : the corresponding spatial frequencies sampling enables a convenient analysis of a complex linear object for a realistic experimental demonstration.}

\end{itemize}
The following sections summarize the THT bench structure. It consists of three main parts (cf Fig.\ref{schema_complet}): a star simulator, a telescope array and a combining interferometer.

\subsection{Star simulator}
\begin{figure}
	\begin{center}
		\includegraphics [width=0.8\linewidth]{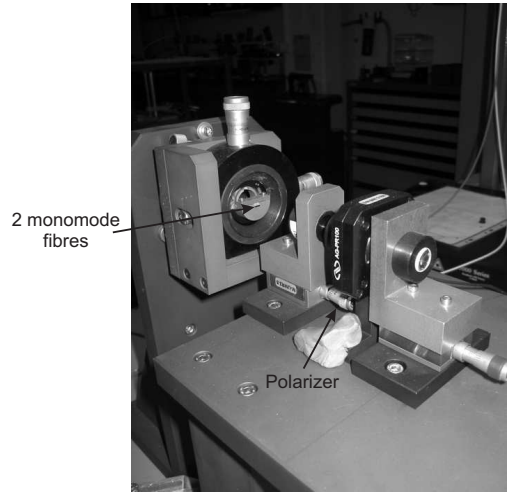}
		\caption{Picture of the experimental star simulator}
		\label{photo_objet}
	\end{center}
\end{figure}

The calibrated object is the first subsystem required for testing the imaging capability of a THT. For this first experimental demonstration, the selected astronomical target is a binary star with a convenient angular separation and adjustable dynamics.\\
For this purpose, the object consists of two tips of  monomode Panda fibres glued on a V-groove. These monomode waveguides are fed by two independent Distributed FeedBack lasers (DFB) with the same emitting wavelength and act as two incoherent point like sources. This way the object is spatially incoherent and the dynamics is controlled by adjusting the laser driving currents. A set of doublets and collimating lenses allows to provide an angular intensity distribution compatible with the spatial frequencies $u$ sampled by our telescope array. In our experiment the angular separation $\theta_{0} $, as seen by the telescope array, is $23.75 \mu rad$. As our instrument is designed for a linear input polarization, a polarizing cube is inserted in the doublet spacing in order to select and fixes a linear vertical input polarization (not drawn on fig \ref{schema_complet}). The experimental setup can be seen in fig.\ref{photo_objet}.

\subsection{Design of the telescope array}
\begin{figure*}
	\begin{center}
		\includegraphics [width=0.8\linewidth]{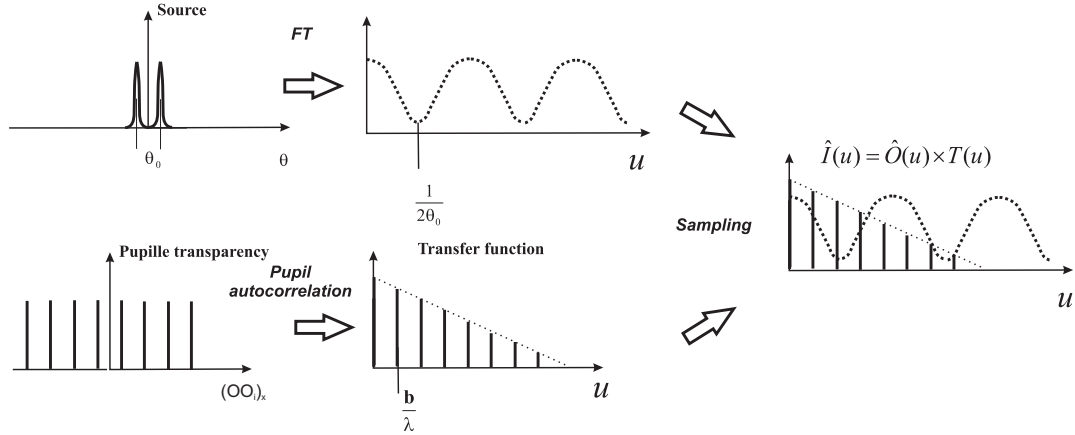}
		\caption{Schematic analysis of the proper sampling criteria. The telescope array design has to be directly related to the angular dimension of the object: $\frac{b}{\lambda }\leq \frac{1}{2\theta _0}$}
		\label{array_design}
	\end{center}
\end{figure*}

The telescope array arrangement has to be carefully selected to fit the sampling criteria for a proper image analysis.
As previously demonstrated \citep{A}, high dynamics imaging capability requires a redundant array configuration. Consequently, our  telescope array must periodically sample  the spatial frequency domain.
The object dimension and the focal length of the collimator have to be determined by comparing the object spectrum and the spatial frequencies sampled by our instrument.
The intensity $I(At;y)$ observed in the image plane of the instrument is given by :
\begin{equation}
I(At;y)=PSF \otimes O(\theta )
\label{equa_PSF}
\end{equation}
where $\otimes $ denotes the pseudo-convolution operator, $PSF$ the point spread function of the instrument and $O(\theta )$, the object intensity distribution.
In the Fourier domain, this relationship becomes a simple product:
\begin{equation}
\hat I(u)=T(u)\times \hat O(u)
\end{equation}
where $\hat O(u)$ is the object intensity spectrum and $T(u)$, the input pupil autocorrelation function.
As shown on Fig. \ref{array_design} the periodicity of the spatial frequencies sampled by the telescope array has to be compliant with the classical Shannon sampling criterium.

\begin{figure}
	\begin{center}
		\includegraphics [width=0.8\linewidth]{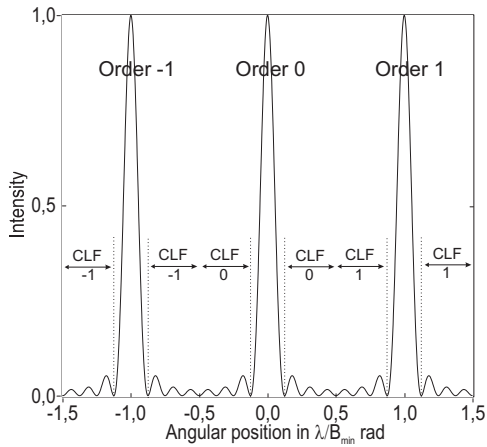}
		\caption{The redundant array configuration induces a periodic PSF with a $\frac{\lambda}{B_{min}}$ pitch.}
		\label{redond_champ}
	\end{center}
\end{figure}

\begin{figure}
	\begin{center}
		\includegraphics [width=0.8\linewidth]{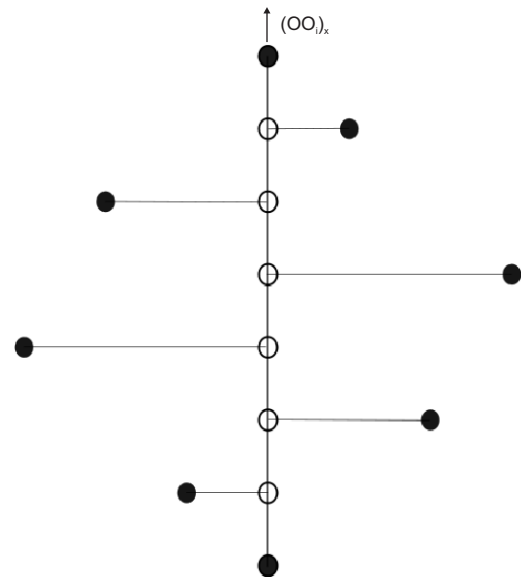}
		\caption{Theoretical (white discs) and experimental (black discs) telescope array configuration: the projection of the telescope baseline along the vertical axis is true to the theoretical redundant configuration. These two configurations are equivalent to a 1D vertical array when used with a 1D vertical object.}
		\label{array_telescope}
	\end{center}
\end{figure}

\begin{figure}
	\begin{center}
		\includegraphics [width=0.6\linewidth]{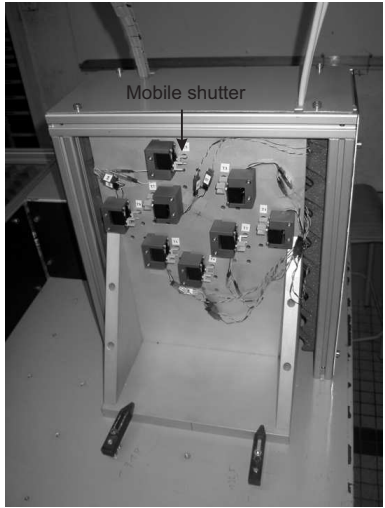}
		\caption{THT input pupil of the telescope array. The lens doublets act as telescopes and mobile shutters allow to adjust the intensity ratio between the interferometric arms in order to get an apodized PSF. }
		\label{photo_reseau}
	\end{center}
\end{figure}

\begin{figure}
	\begin{center}
		\includegraphics [width=\linewidth]{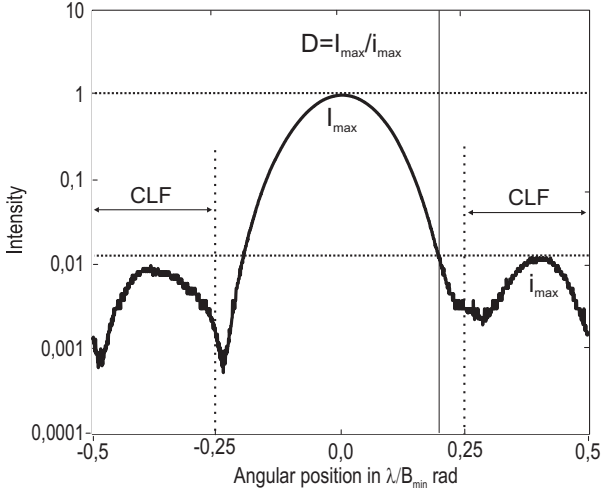}
		\caption{PSF dynamics $D$ is defined as the ratio between the maximum intensity on the PSF $I_{max}$ and the higher intensity over the Clear Field $i_{max}$}
		\label{def_dynamique}
	\end{center}
\end{figure}

The smallest sampled frequency ($\frac{B_{min}}{\lambda}$) is related to the instrument field of view $FV$ and the largest one ($\frac{B_{max}}{\lambda}$) determines  the instrument resolution $R$:
\begin{equation}
FV=\lambda/B_{min}
\label{FV}
\end{equation}
\begin{equation}
R=\lambda/B_{max}
\end{equation}
The telescope array resolution allows to discriminate the sharpest details to be observed on the object, and the field of view has to be adapted to the object's overall size.\\
These design trade-offs lead to the following THT bench characteristics:
$$FV=\frac{\lambda}{B_{min}}=62 \mu rad$$
$$R=\frac{\lambda}{B_{max}}=8.9 \mu rad$$
These characteristics have been chosen to be compatible with the observation of our laboratory binary star characterized by a $23,75 \mu rad$ angular separation.
$$R < \theta_{0} < FV$$ 
Notice that a redundant array, as proposed in this paper, induces a periodic PSF as reported on figure \ref{redond_champ}.  The equation \ref{FV}  gives the Field of View limits proper to avoid aliasing around the 0 interferometric order. The area between two consecutive main lobes is called Clear Field (cf Fig.\ref{redond_champ}). In the case of polychromatic sources, only the zero interferometric order is achromatic and consequently appropriate for direct imaging.
During the implementation of our breadboard we faced compactness constraint due to the mechanical dimension of the launching assemblies. The focusing lenses are small enough to allow a compact linear array but the 3 axes nanopositioning mountings, required for the fine adjustment of the fibre inputs, cannot be aligned with a pitch smaller than 25 mm. So far, the real THT telescope array has been designed in a two-dimensional configuration but dedicated to observe a one dimensional object along the vertical axis. The projection of the telescope positions on the vertical axis is true to the expected linear array design as shown in Fig.\ref{array_telescope}. In such a configuration, the instrumental response is exactly the same as with the linear telescope array when observing a vertical linear object.\\
In order to image an unbalanced binary system with very high dynamics, the use of a set of suitable apodization coefficients will optimize the PSF dynamics  with a low decay of the resolution (cf Fig \ref{psf_ideal}). On each aperture, the  control of the intensity  is achieved by means of mobile shutters (cf Fig.\ref{photo_reseau}) actuated with a high position accuracy according to the theoretical optimum distribution \citep{A}. PSF dynamics $D$ is defined as the ratio between the maximum intensity on the PSF $I_{max}$ and the higher intensity over the Clear Field $i_{max}$ (cf fig\ref{def_dynamique}).

\subsection{The interferometeric combiner}
\begin{figure}
	\begin{center}
		\includegraphics [width=\linewidth]{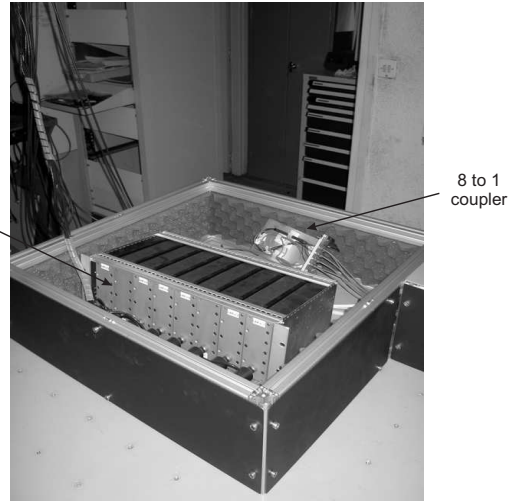}
		\caption{The THT interferometric beam combiner including the optical fibre delay lines and the optical path modulators.}
		\label{photo_interferometre}
	\end{center}
\end{figure}

The last part of the system is the optical field combiner mixing the contributions of the 8 telescopes (cf Fig\ref{photo_interferometre}). Each interferometric arm includes a fibre delay line and an optical path modulator \citep{O}. The 8- fibre arms have been cut with a few mm accuracy in order to reduce the optical path differences as much as possible. The fibre delay lines allow to adjust the optical path with an accuracy of few $\mu m$. The fibre optical path modulators temporally reproduce the linear phase variation observed in the image plane of "classical" hypertelescopes (cf eq.\ref{phase_equa}) and allow the fine residual optical path compensation. A National Instrument virtual instrument and a voltage generator have been implemented and developed to drive the piezoelectric actuators of the fibre optical path modulators. A set of high voltage amplifiers allows to reach a proper range of command voltages. Experimentally, the optical path control is achieved with an optical path sensitivity over $0.01 \mu m$.\\
The  optical fields emerging from the 8 interferometric arms reach a 8 to one polarization maintaining (PM) coupler to achieve the interferometric mixing. At the output an InGaAs photodiode  detects the interferometric signal that is recorded through a standard 12 bits ADC voltage acquisition system.

\section{ Experimental test and imaging capabilities of the THT test-bench}
\subsection{PSF experimental results}
\begin{figure}
	\begin{center}
		\includegraphics [width=0.8\linewidth]{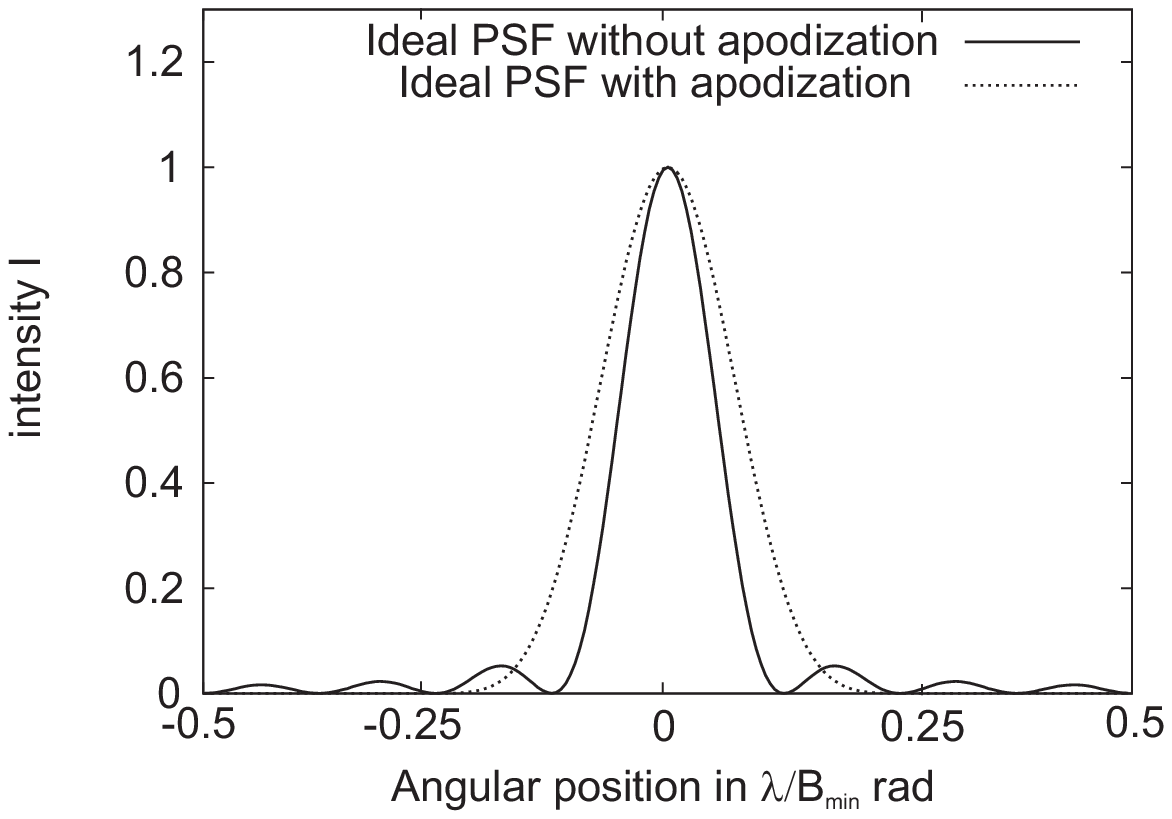}
		\includegraphics [width=0.8\linewidth]{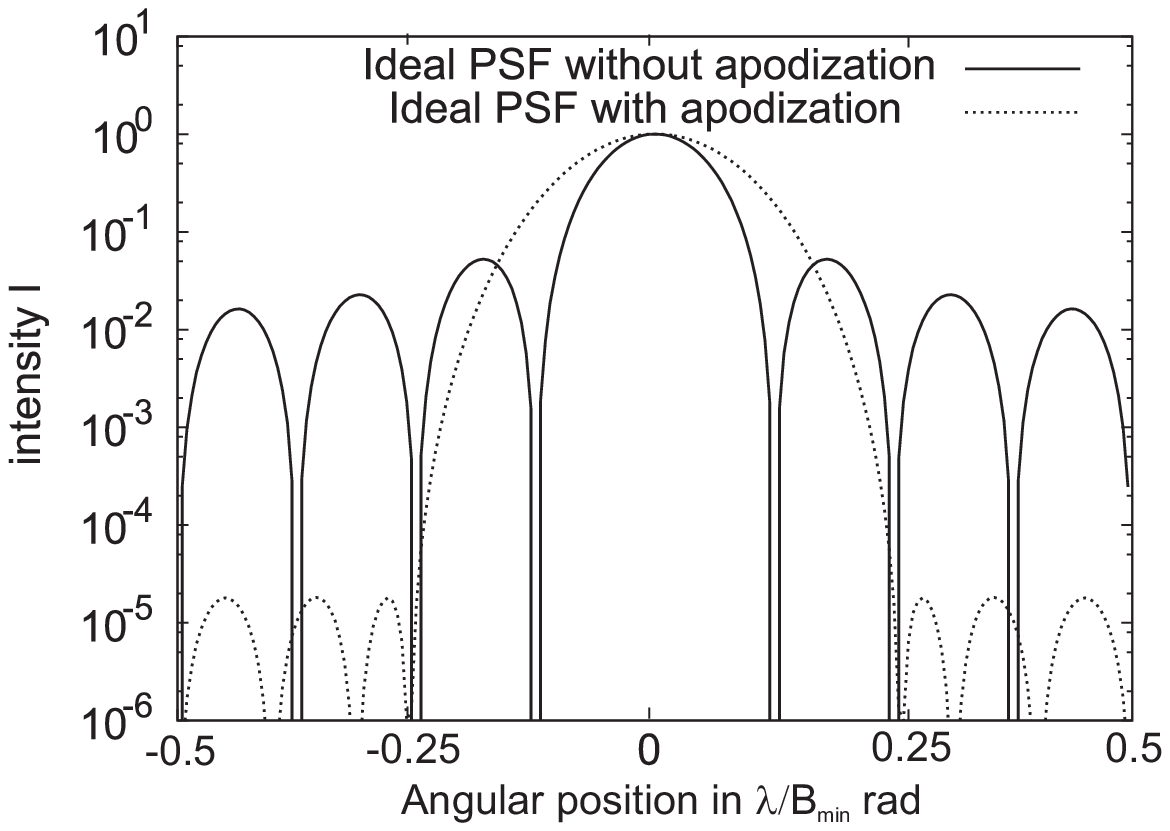}
		\caption{Theoretical influence of the apodization process on the instrument PSF: the apodization allows to observe high dynamics objects with a relatively low decay of the resolution (top linear scale, bottom log scale).}
		\label{psf_ideal}
	\end{center}
\end{figure}
\begin{figure}
	\begin{center}
		\includegraphics [width=0.8\linewidth]{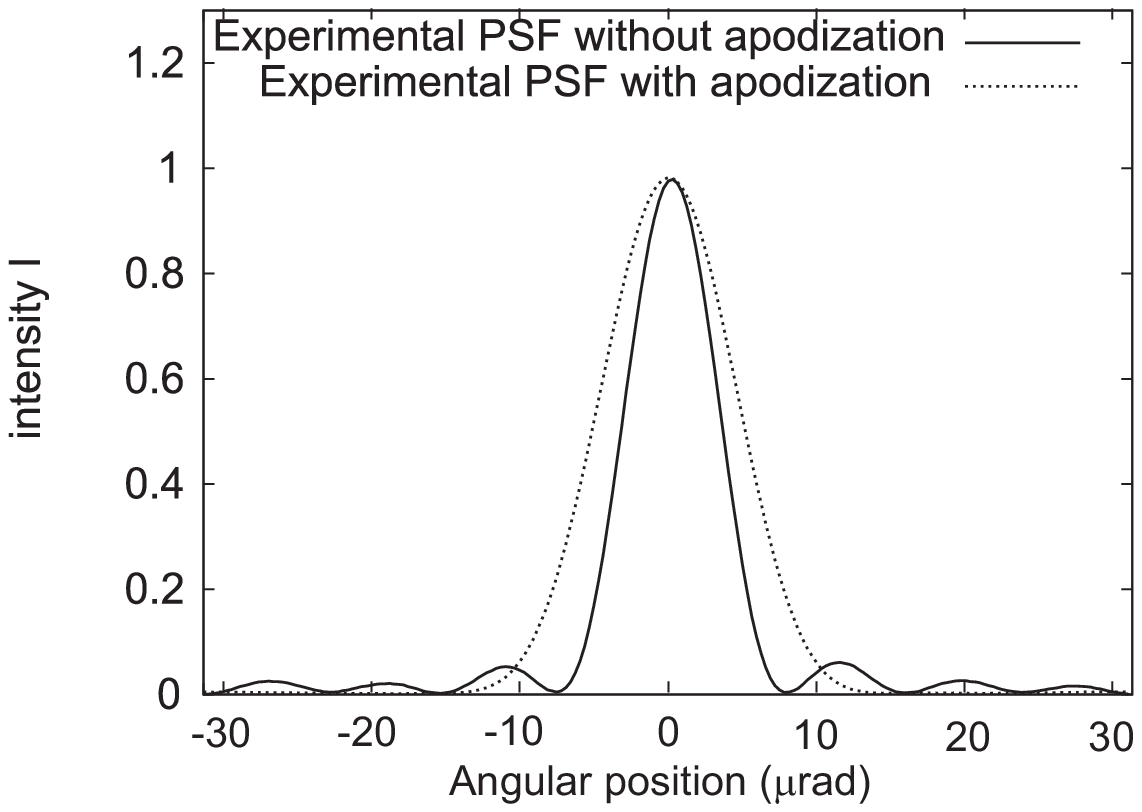}
		\includegraphics [width=0.8\linewidth]{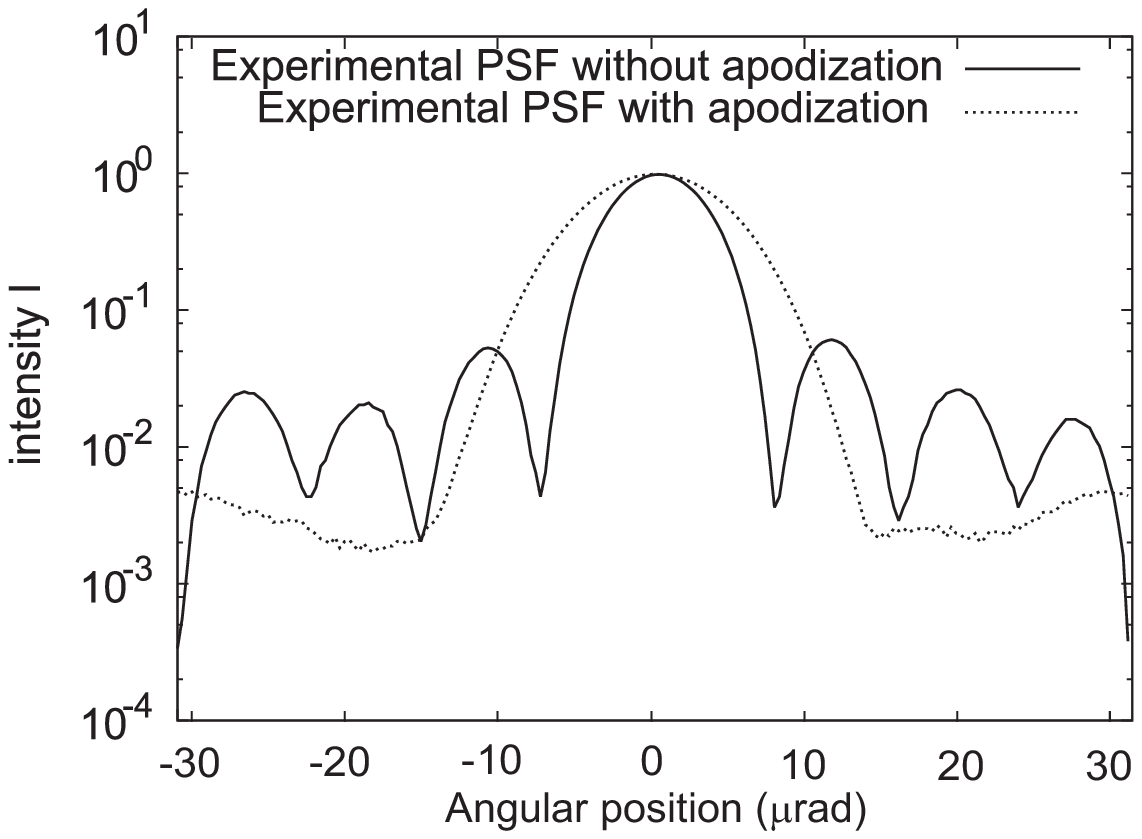}
		\caption{Experimental PSF obtained on our 8-telescope THT test bench: The apodization process allows to reach a 300 dynamics between the maximum and the clear field area.(top linear scale, bottom log scale)}
		\label{experimental_PSF}
	\end{center}
\end{figure}
\begin{figure}
	\begin{center}
		\includegraphics [width=1\linewidth]{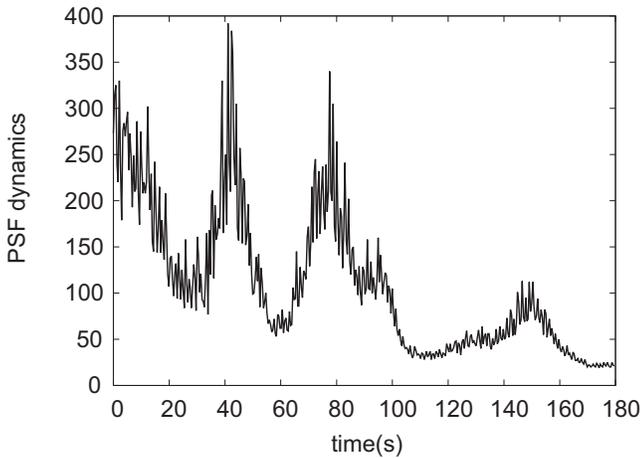}
		\caption{PSF dynamics as function of time: example of the evolution of experimental PSF dynamics  over 180 s.}
		\label{dyn_evol}
	\end{center}
\end{figure}
In order to calibrate the imaging properties of our instrument, the first investigation has been to characterize its point spread function. For this purpose, the telescope array has been illuminated by a plane wave using only a single point-like source ( i.e. switching on only 1 fibre of the object). A servo control system not available during this experiment, the 8 telescope cophasing has been achieved manually taking advantage of the relative stability of our instrument. The voltage offsets of the electronic commands, driving the optical path modulators, have been adjusted in order to increase the dynamics as well as possible. This is possible if you carrefully avoid any perturbation of the instrument over the cophasing process. For this purpose, the instrument has been acoustically and thermally baffled. The first experimental results are shown in  Fig.\ref{experimental_PSF}. They are consistent with the theoretical PSF simulation proposed in Fig.\ref{psf_ideal}. The second main feature to be analyzed is the influence of the apodization process  as shown on fig \ref{experimental_PSF}. Using the apodozation coefficient computed in ref \citep{A} allows to reduce consequentially the ripple in the clear field all around the PSF peak. The best dynamics experimentally observed with our breadboard are in the range of 300 but as shown on figure \ref{dyn_evol} these results are not stable over time, probably due to environmental perturbations such as acoustic waves, vibrations or temperature fluctuations. Nevertheless, such results have been obtained reliably and give a first indication on the THT imaging potential capabilities. As we will demonstrate in the next paragraph, the major limitation on the dynamics results from the manual cophasing that remains of poor quality when compared with a servo controlled system. The main effect is a reduction of the dynamics that is particularly clear in the apodized configuration.  These first tests demonstrate the possibility to image a point-like source on axis.

\subsection{Analysis of experimental results}
The phenomena able to reduce the PSF dynamics can be listed as follows:
\begin{itemize}
\item{The residual optical path difference between the 8 interferometric arms.}
\item{The differential polarization properties between the 8 fields to be mixed due to differential birefringent properties of the interferometric arms.}
\item{The difference between the theoretical results reported in \citet{A} and the apodization really applied in the experiment.}
\item{The differential dispersion properties between the 8 interferometric arms.}

\end{itemize}
This last point is not relevant in our experiment due to the use of a quasi-monochromatic source.\\
\begin{figure}
    \begin{center}
          \includegraphics [width=.8\linewidth]{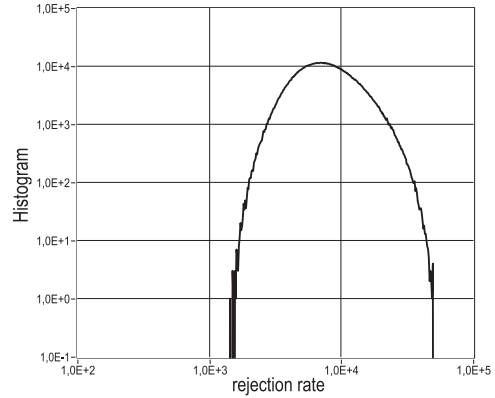}
     \caption{Histogram on minimum dynamics as function of fluctuations on the apodization coefficient: this histogram is obtained from one million of simulated PSF. For each PSF, the apodization coefficient values are determined randomly using the experimental statistic ( mean value and RMS). This numerical simulation demonstrates that in the worst case of 1500 dynamics is far better than the experimental result. Accordingly, this error source has a minor effect.}
      \label{erreur_apod}
    \end{center}
 \end{figure}
The influence of the third point has been analyzed by modeling the effect of the experimental uncertainties on apodization coefficients. According to the error experimentally observed, the $a_i$ coefficients are randomly changed in a numerical simulation and the final PSF is computed. Table \ref {tableau_apodisation} summarizes the experimental uncertainties observed in our test bench. This statistical distribution on $a_i$ coefficient has been used as input for our numerical simulations. A set of PSF has been computed varying the $a_i$ allowing to plot a histogram of the rejection ratio ( Fig. \ref{erreur_apod}). This curve demonstrates that the probability to get dynamics lower than 1500 is practically zero. This dynamics is far higher than the observed value. Accordingly, the experimental uncertainty on the apodization coefficient cannot explain the observed experimental rejection ratio and cannot be identified as responsible for the 300 experimental dynamics.

\begin{table}
\begin{center}
\begin{tabular}{|*{4}{c|}}
\hline
Arm & apodization & theoretical & experimental\\
 i & coeff & normalized & intensity\\
 & $a_i$ & intensity & range\\
\hline
1 - 8 & 0.104 & 0.011 & [0.005;0.016]\\
\hline
2 - 7 & 0.364 & 0.133& [0.122;0.143]\\
\hline
3 - 6 & 0.724& 0.525& [0.496;0.553]\\
\hline
4 - 5 & 1 & 1.000 & [0.985;1.015]\\
\hline
\end{tabular}
\caption{Apodization parameters}
\label{tableau_apodisation}
\end{center}
\end{table}

\begin{figure}
    \begin{center}
          \includegraphics [width=.8\linewidth]{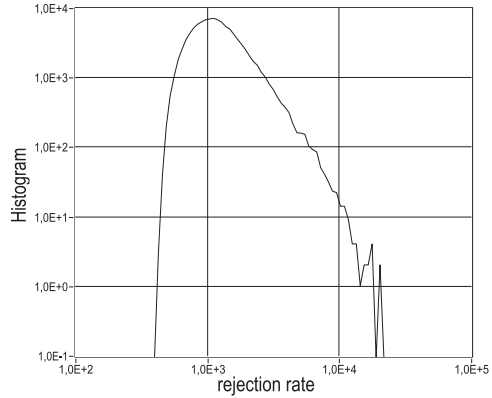}
     \caption{Birefringence rejection ratio limitation: This histogram obtain by 100000 PSF simulation shows the birefringence effect on the rejection ratio in the case of our experimental setup (contrast = 0.999). In the worst case, the rejection ratio is limited to 500.}
     \label{erreur_contrast}
    \end{center}
 \end{figure}

In order to carry on our investigation on with the experimental origin of the limited dynamics, the next point under study is the polarization behaviour.
The optical fibre arms of the interferometer are made of polarization-maintaining monomode fibres. This choice results from the necessity to avoid any loss of the polarization coherence between the optical fields to be interferometrically mixed. There can be two problems :
\begin{itemize}
\item{Input linear polarization can be misaligned with the used neutral axis at the input of waveguides.}
\item{Crosstalk between the two polarization modes corresponding to the two neutral axes.}
\end{itemize}    
 In both cases, it results in an incoherent background intensity propagating along the undesirable polarization axis. This parasitic intensity leads to a background limit reducing the dynamics to be reached by the instrument. In our experiment, care has been taken to limit this problem by means of polarizers placed at the entrance pupil and after the 8 to 1 optical coupler. Our bread board has been characterized by measuring the contrast with each telescope couple. All the contrasts remain better than 0.999 as reported in Table.\ref{tableau1}. According to these experimental data, it is possible to compute a global PSF introducing a random polarization crosstalk for each interferometric arm. It results in a histogram of the rejection ratio to be reached according to such a perturbation (Fig.\ref{erreur_contrast}). This simulation shows that the probability to get a rejection ratio lower than 500 is practically zero. Accordingly, the polarization behaviour cannot be identified as responsible for the 300 dynamics observed in our experiment.
Consequently, the 300 dynamics can only be explained by the residual optical path differential fluctuations between the 8 interferometric arms.
\begin{table}
\begin{center}
\begin{tabular}{|*{2}{c|}}
\hline
spatial & experimental\\
frequencies&contrast\\
\hline
1 & 0.9996 ($\pm 4.10^{-4}$) \\
\hline
2 & 0.9995 ($\pm 4.10^{-4}$)\\
\hline
3 & 0.9993 ($\pm 3.10^{-4}$)\\
\hline
4 & 0.9994 ($\pm 4.10^{-4}$)\\
\hline
5 & 0.9992 ($\pm 3.10^{-4}$) \\
\hline
6 & 0.9992 ($\pm 3.10^{-4}$)\\
\hline
7 & 0.9991 ($\pm 3.10^{-4}$)\\
\hline
\end{tabular}
\end{center}
\caption{Experimental contrasts as function of the sampled spatial frequencies achieved with pairs of telescope.}
\label{tableau1}
\end{table}

\subsection{THT imaging capabilities}
\begin{figure}
	\begin{center}
		\includegraphics [width=0.8\linewidth]{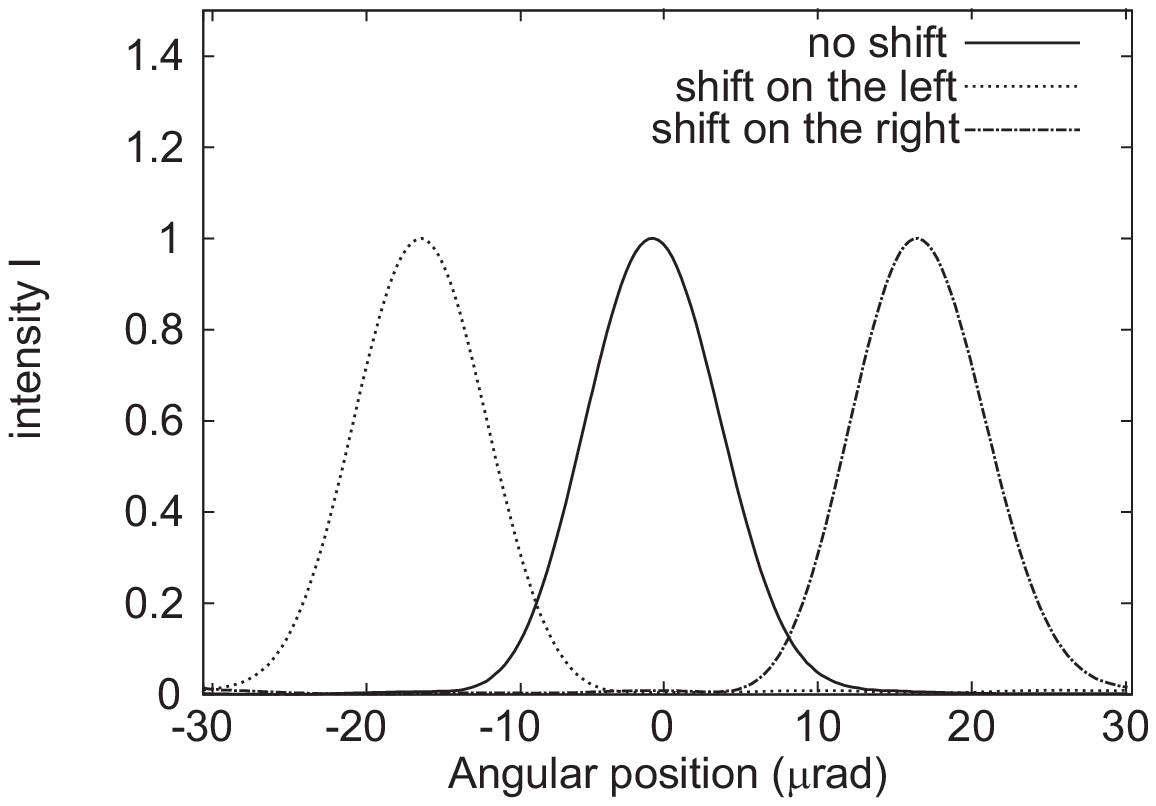}
		\includegraphics [width=0.8\linewidth]{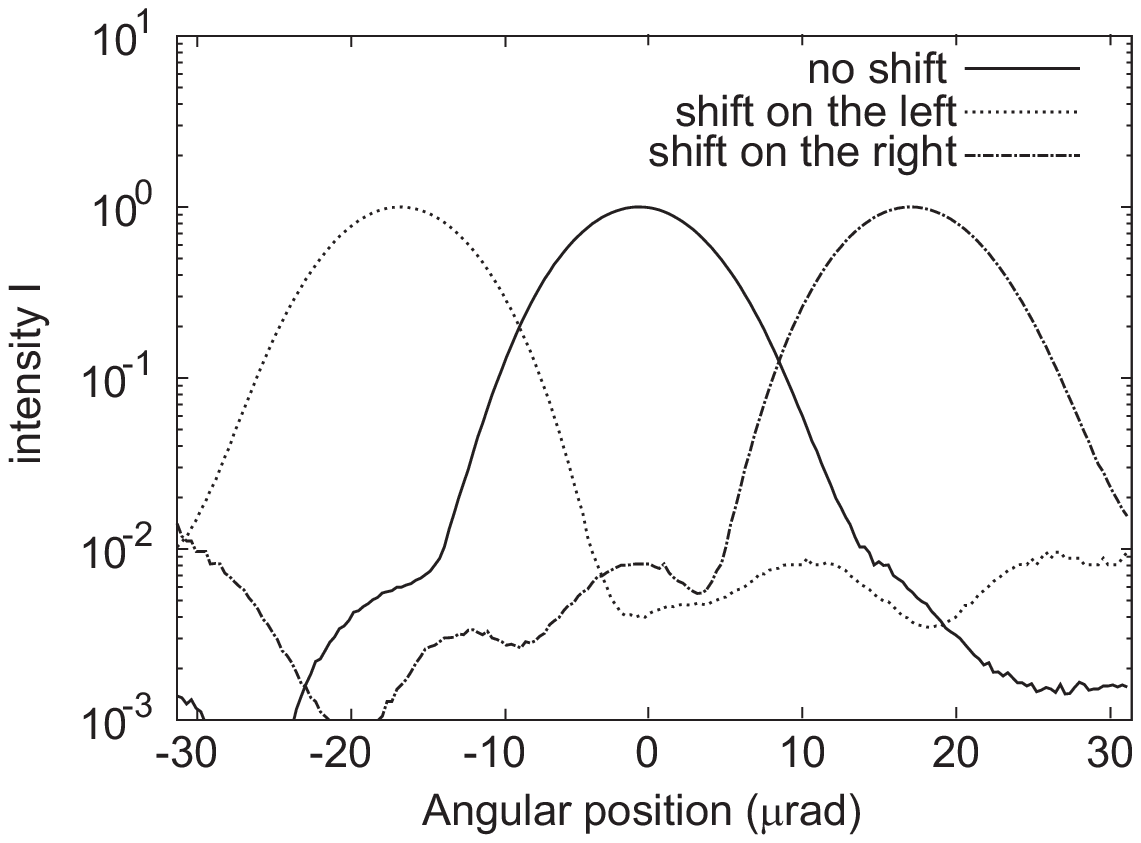}
		\caption{PSF translation quasi invariance: Moving the point-like source in THT field of view allows to observe global shift of the PSF.The repeatability is experimentally demonstrated only at a 100:1 level due to the turbulence phase fluctuation.}
		\label{psf_shift}
	\end{center}
\end{figure}
\begin{figure}
	\begin{center}
		\includegraphics [width=0.8\linewidth]{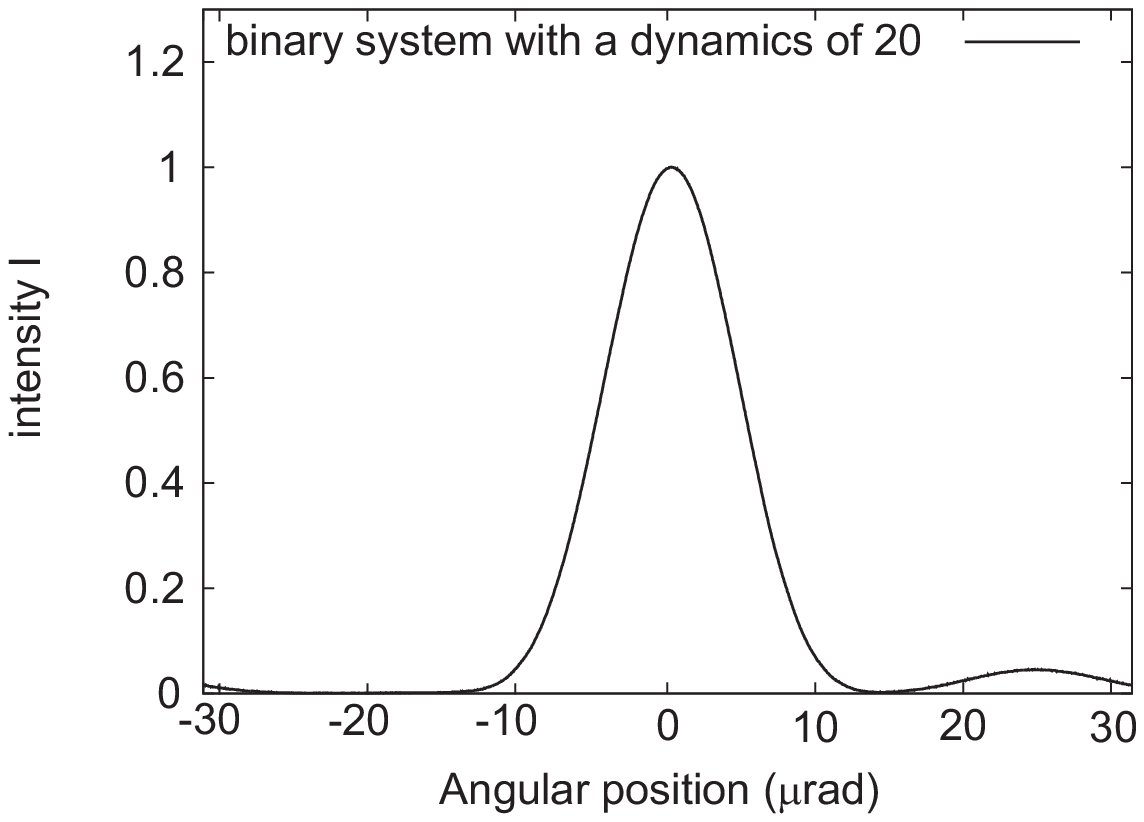}
		\includegraphics [width=0.8\linewidth]{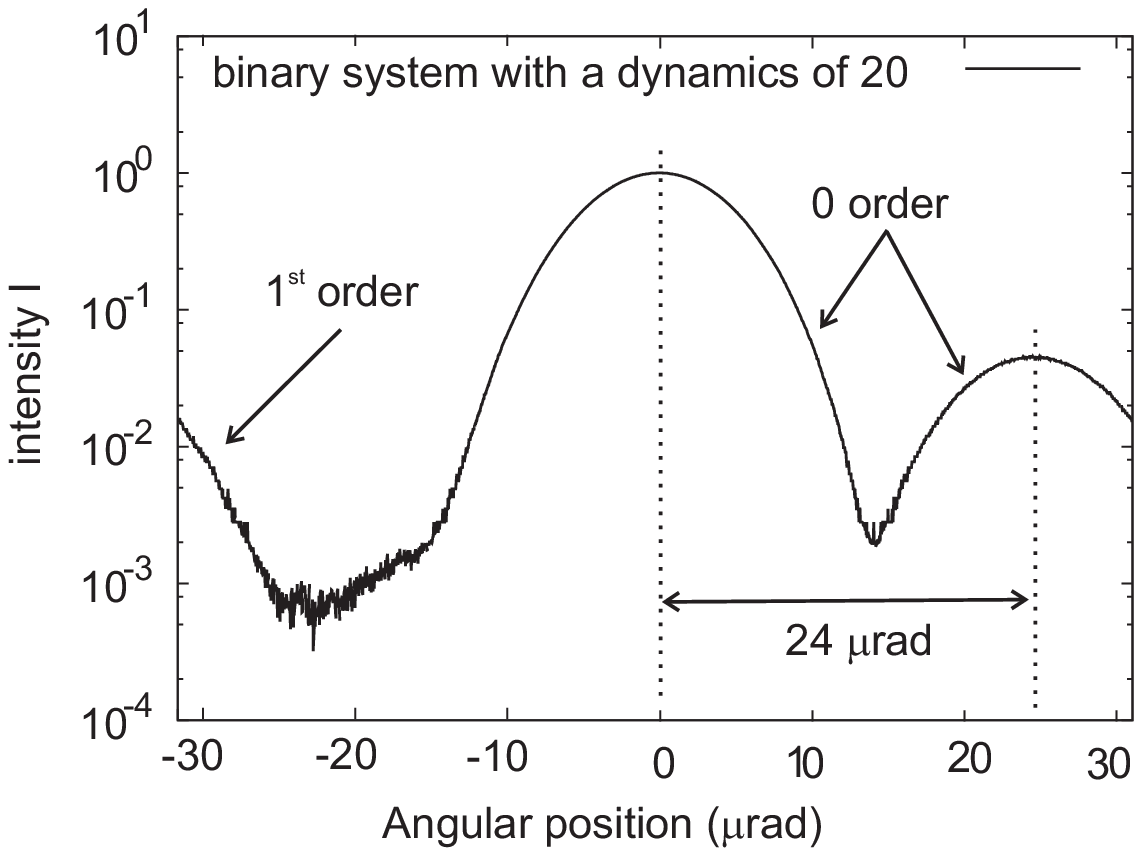}
		\caption{Experimental image of an unbalanced binary system with a 20 power ratio using our 8-telescope array in the apodized configuration(top linear scale, bottom log scale).}
		\label{binaire}
	\end{center}
\end{figure}

The second stage of our experimental investigation has been to check the invariance of this point spread function over the imaging field of view. As a preliminary test, we moved the point like source in the object plane by shifting the optical fibre tip using manual translations. As shown on figure \ref{psf_shift}, the PSF was simply shifted with a constant width of the main lobe. The intensity on the clear field shows some fluctuations leading to dynamics losses. These imperfections are due to environmental parameter fluctuations resulting from the lack of a real time cophasing system.
This demonstrates that our instrument is linear and translation invariant and therefore suitable for imaging .\\
In order to illustrate the imaging capabilities of our instrument, we have used a binary star as source. The two fibre tips of the laboratory object receive the light from two different lasers operating at the same wavelength. Adjusting the driving current allows to simulate any unbalanced binary source. In the absence of a real time cophasing system, we used a secondary star intensity higher than the clear field intensity fluctuations.

The experimental result is shown in fig.\ref{binaire} and exhibits a main lobe surrounded by two side lobes. The bright one is related to the main star and the two others to its companion with the 0 interference order on the right and the first one on the left of this figure. Notice that in the case of polychromatic sources, only the zero interferometric order will be used as being achromatic. The source characteristics with a $23.75 \mu rad$ separation  and a $\sim 20$ intensity ratio are correctly recovered in the image.
Results are consistent with the theoretical convolution between the point spread function and the object intensity distribution according to eq.\ref{equa_PSF}. Consequently, over the instrument clear field the instrument imaging capabilities have been demonstrated.

\subsection{Preliminary 2D PSF observation}
\begin{figure}
    \begin{center}        
	 \includegraphics [width=.8\linewidth]{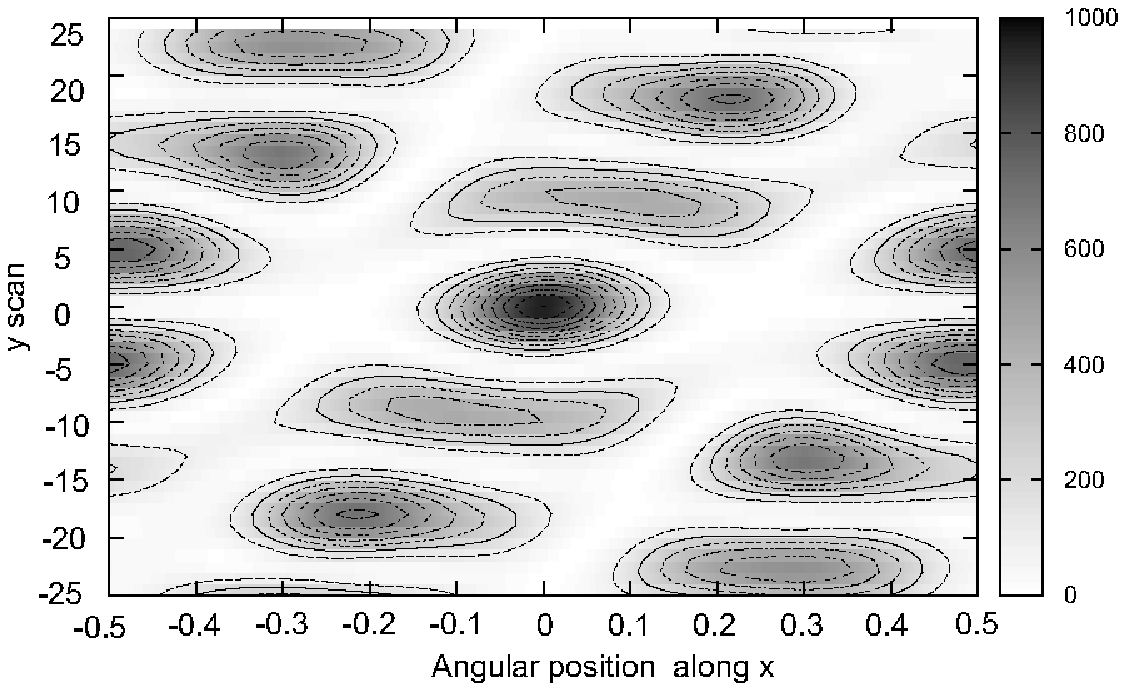}
     \includegraphics [width=.8\linewidth]{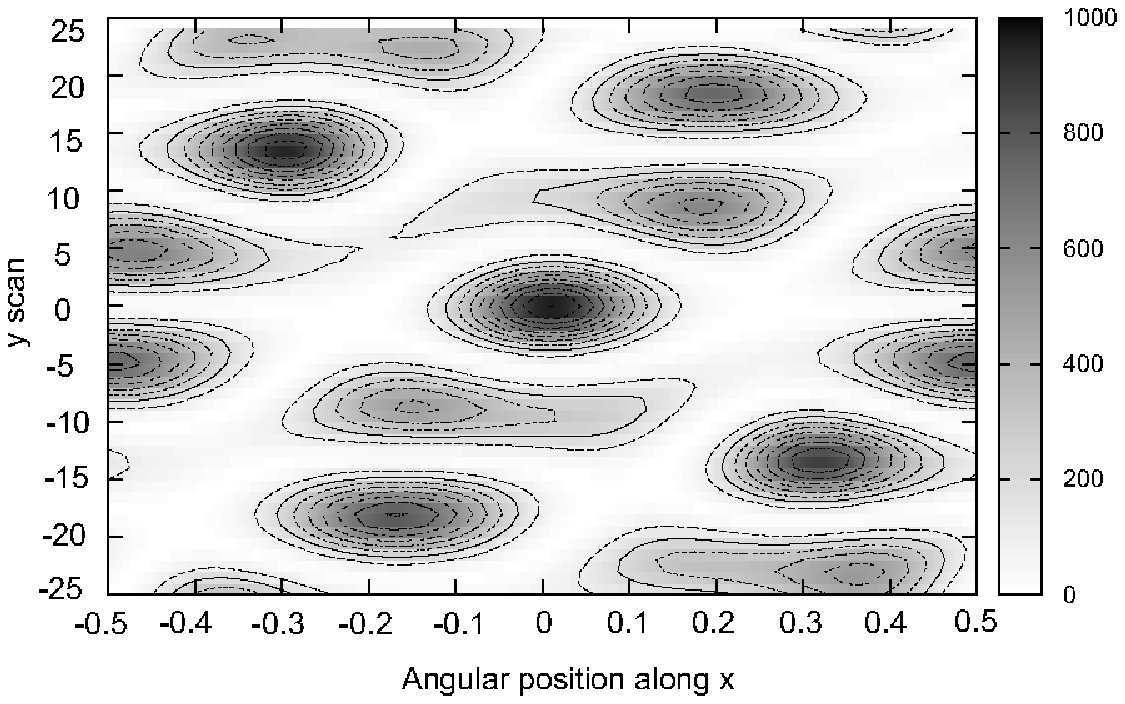}
     \includegraphics [width=.8\linewidth]{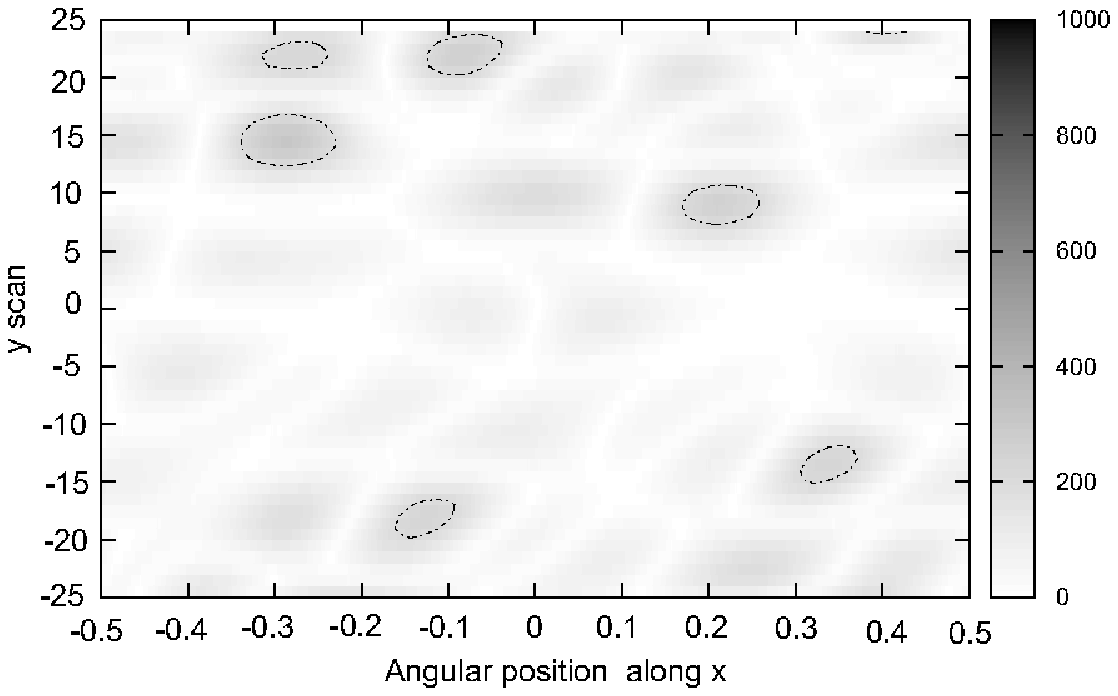}
     \caption{Preliminary test on the 2D imaging capabilities of a Temporal Hypertelescope. Comparison between ideal (top) and experimental (middle) 2D PSF obtained with our THT bench (bottom)absolute value of the difference between the ideal PSF and the experimental image.}
     \label{psf_2D}
    \end{center}
 \end{figure}

This first experimental study was mainly dedicated to the demonstration of the imaging capabilities of a temporal hypertelescope on a one dimensional object. Due to mechanical constraints, the telescope array has been designed in a 2D configuration. Even if this array is not absolutely optimized for a two dimensional imaging, it is possible to test the 2D PSF to demonstrate the operation of the 2D imaging process.  This picture can be obtained row by row like  in a TV raster scan using the phase modulations reported in equation eq.\ref{phase_equa_temporal}. A set of phase shifts $\varphi_{i0}$ is sent through the driving voltage electronics taking into account the projection of the telescope  $i$ baseline along the $y$ direction :
\begin{equation}
\varphi_{i0}(y_0)=2\pi.\beta_i.y_0
\end{equation}
This equation describes the phase shift that could be observed  for telescope $i$ in the image plane along the vertical axis in a spatial configuration.
As demonstrated in ref (\citet{RD}) it is possible this way to get a 2D information using a temporal hypertelescope. We tested this process by recording the 2D PSF using our experimental test bench. The theoretical and experimental patterns are reported on Fig.\ref{psf_2D}.
The array cophasing was manually achieved during the first scan on the central row ($y_0=0$). After this initialization process, it is possible to scan all the field of view taking advantage of the relative stability of the telescope array. This full operation lasts about 20 seconds to get 50 rows over the 2D PSF. In order to quantify the discrepancies between ideal (Figure \ref{psf_2D} top) and experimental (Figure \ref{psf_2D} middle) 2D PSF, Figure \ref{psf_2D} bottom reports the absolute value of the difference between the two images. The significant fluctuations are probably due to the lack of an active phase control system during the 2D scan. A fine study of this limitation will be achieved after the implementation of a servo control system of our test bench.  Nevertheless, these results are promising and could be enhanced with a convenient cophasing system. 
\section{Conclusion}
The aim of this study was to experimentally demonstrate the operation of a temporal hypertelescope. This experimental demonstration was achieved by testing the instrument Point Spread Function as a first stage. The second experiment demonstrated the possibility to observe an unbalanced binary star with a 20 flux ratio. During this first test, the PSF dynamics was limited to 300  due to the lack of servo control system. The next stage of this study will be the design and implementation of a servo control system to accurately cophase our telescope array in order to enhance the dynamics of our instrument.
\section*{Acknowledgements}
This study has been financially supported by CNES, INSU and Thales Alenia Space in the frame of different contracts.
Our thanks go to Emmanuelle Abbott for her help in writing this paper and Alain Dexet for the fabrication of the mechanical parts of our experiment.

\appendix
\bsp
\label{lastpage}
\end{document}